\documentclass[copyright,creativecommons]{eptcs}
\providecommand{\event}{CL&C 2010}
\usepackage{amsmath}
\usepackage{amssymb}
\usepackage{mathrsfs}
\usepackage{bussproofs}
\usepackage{upgreek}
\usepackage{microtype}
\usepackage{hyperref}

\usepackage{pdfsync} 

\usepackage{mathrsfs}

\title{Strong Normalization for HA + EM1 by Non-Deterministic Choice
}
\author{Federico Aschieri\footnote{This work was supported by the LABEX MILYON (ANR-10-LABX-0070) of Universit\'e de Lyon, within the program ``Investissements d'Avenir'' (ANR-11-IDEX-0007) operated by the French National Research Agency (ANR)}\institute{Laboratoire de l'Informatique du Parall\'elisme (UMR 5668),  \'equipe Plume\\
\'Ecole Normale Sup\'erieure de Lyon -- Universit\'e de Lyon\\
France
}}

\def\titlerunning{Strong Normalization for $\HA + \EM_{1}$ by Non-Deterministic Choice}
\def\authorrunning{F. Aschieri}

\newcommand{\comment}[1]{}
\newcommand{\EM}                       { {\mathsf{EM}} }
\newcommand{\NEM}                       { {\mathsf{EM_{1}^{\star}}} }
\newcommand{\HA}                       { {\mathsf{HA}} }
\newcommand{\Language}                 {\mathcal{L}}
\newcommand{\E}[3]                   {{ #2 \parallel_{#1} #3}}

\newcommand{\inj}                   {{{\upiota}}}
\newcommand{\emp}[1]    {{\mathsf{P}_{#1}}}
\newcommand{\Hyp}[2]                   {{\mathtt{H}^{\forall {#2} \mathsf{#1}}}}
\newcommand{\wit}                 {{\mathtt{W}_{\emp{}}}}
\newcommand{\Wit}[2]               {{\mathtt{W}^{\exists {#2} \lnot\mathsf{#1}}}}

\newcommand{\red}              {\;\mathsf{r}\;}
\newcommand{\cruno}           {{\textbf{(CR1)}}}
\newcommand{\crdue}           {{\textbf{(CR2)}}}
\newcommand{\crtre}           {{\textbf{(CR3)}}}
\newcommand{\crquattro}           {{\textbf{(CR4)}}}
\newcommand{\num}[1]{\overline{#1}}
\newcommand{\nSystemT}{\mathsf{T}^{\star}}

\newcommand{\SystemTG}                  {\mathsf{T}}
\newcommand{\ifn}{{\mathsf{if}}}

\newcommand{\Nat}                      { {\tt N} }
\newcommand{\Bool}                     { {\tt Bool} }
\newcommand{\sn}{\mathsf{SN}}
\newcommand{\rec}                          {{\mathsf{R}}}

\newcommand{\True}                     { {\tt{True}} }
\newcommand{\False}                    { {\tt{False}} }
\newcommand{\suc}{\mathsf{S}}
\newcommand{\itr}{\mathsf{It}}
\newcommand{\eq}             {{\mathsf{eq}}}
\newcommand{\add}             {{\mathsf{add}}}
\newcommand{\mult}             {{\mathsf{mult}}}

\newcommand{\pair}[2]{\langle #1,#2\rangle}
\newcommand{\prj}[2]{\pi_{#1}{[#2]}}

\newcommand{\trans}[1] {{#1}^{*}}

\newcommand{\redn}              {\rightsquigarrow}
\newcommand{\nf}{\mathsf{NF}}

\newcommand{\redcbv} {\mapsto_{\mathsf{cbv}}}

\newcommand{\dlinea}{\leavevmode\hrule\vspace{1pt}\hrule\mbox{}}

\newtheorem{theorem}{Theorem}
\newtheorem{corollary}{Corollary}

\newtheorem{proposition}{Proposition}
\newtheorem{definition}{Definition}

\begin{document}
\maketitle

\begin{abstract}
We study the strong normalization of a new Curry-Howard correspondence for  $\HA + \EM_{1}$, constructive Heyting Arithmetic with the excluded middle on $\Sigma^{0}_{1}$-formulas. The proof-term language of $\HA+\EM_{1}$ consists in the lambda calculus plus an operator $\E{a}{}{}$ which represents, from the viewpoint of programming, an exception operator with a delimited scope, and from the viewpoint of logic, a restricted version of the excluded middle. We give a strong normalization proof for the system based on a technique of ``non-deterministic immersion''.
\end{abstract}

\section{Introduction}

In the field of computer science and proof theory that studies the \emph{classical} Curry-Howard correspondence \cite{Sorensen} between proofs and programs, there are essentially two ways of showing that it is possible to extract useful information from deductions and their associated programs. 

The first approach is quite old, and belongs to the tradition of proof theory. It consists in proving \emph{separately}  strong normalization results -- the execution of all proof terms always terminates -- and normal form properties -- all the proof terms which terminate  in normal form have some particular well-characterized shape. Examples of this approach may be found in Prawitz \cite{Prawitznatural}, the first to have defined reduction rules for classical natural deduction proofs (see also Barbanera and Berardi \cite{BB1}); or in Barbanera and Berardi \cite{BB2,BB3}, for classical Arithmetic with control operators or symmetric lambda calculus; or in Parigot \cite{Parigot1,Parigot2}, in the case of the  $\lambda\mu$-calculus for second-order logic (many other examples can be found in the literature). 

The second approach consists in defining a realizability relation for classical logic, and use it as a tool to deduce properties of proof terms. While this approach worked very well for intuitionistic logic (see e.g. Troelstra \cite{Troesltra} and Krivine \cite{Krivine1}), it took quite a while to adapt it to classical logic and classical Arithmetic. Examples of this approach may be found in Krivine \cite{Krivine2,Krivine3}, where it is defined a realizability for second order classical logic and even set theory; or in Avigad  \cite{AvigadR}, where it is proposed a realizability for classical Arithmetic akin to Coquand's game semantics; or in Aschieri and Berardi \cite{ABF},  Berardi and de' Liguoro \cite{BerardiLiguoroMonadi}, Aschieri \cite{AschieriCSL} where it is introduced Interactive realizability for classical Arithmetic, even with first-order choice axioms. 

The two approaches are of course related: many proofs of strong normalization use Tait-Girard reducibility techniques \cite{Girard}, which can often be seen as  special cases of realizability. But especially for predicative systems of classical Arithmetic, there are many normalization proofs available, which have nothing to share with realizability.  

This paper falls in the first category of contributions: our goal is to prove the strong normalization of a new set of reduction rules for Heyting Arithmetic $\HA$ with the excluded middle schema $\EM_{1}$, $\forall \alpha^{\Nat}\emp{}\lor \exists \alpha^{\Nat} \lnot\emp{}$, where $\emp{}$ is any atomic decidable predicate.   That Curry-Howard correspondence for $\HA+\EM_{1}$ has been presented in \cite{ABB,Birolo} and it is based on delimited exceptions, and permutative conversions for $\EM_{1}$-disjunction elimination. Permutative rules were introduced by Prawitz (see \cite{Prawitz}) to obtain the subformula property in first-order natural deductions. Delimited exceptions were used by de Groote \cite{deGrooteex} in order to interpret the excluded middle in classical propositional logic with implication; by Herbelin \cite{Herbelin}, in order to pass witnesses to some existential formula when a falsification of its negation is encountered: in our setting they are used in a similar way, and our work may be seen as a modification and extension of some of de Groote's and Herbelin's techniques. Many of our ideas are inspired by Interactive realizability \cite{ABF} for $\HA+\EM_{1}$, which describes classical programs as programs that makes hypotheses, test them and learn by refuting the incorrect ones. 

In \cite{ABB},  the realizability approach has been used to study the system  $\HA+\EM_{1}$; the main soundness theorem of realizability entails strong normalization and the witness extraction property, i.e. the possibility of computing witnesses for simple existential statements.  In \cite{Birolo}, instead, it has been presented a syntactical proof of a normal form property for $\HA+\EM_{1}$, which gives as corollary the witness extraction property. Here, we provide the missing piece to complete an approach of the first category: we give a direct proof of strong normalization for $\HA+\EM_{1}$. 

Our main ideas are new, and are neither based on some continuation-passing-style translation as in countless  proofs of classical  strong normalization (see e.g. Griffin \cite{Griffin} or de Groote \cite{deGroote}) nor on a Parigot-style reducibility argument (see e.g. \cite{Parigot2,Nour}). Indeed, both these last approaches correspond, from a logical point of view, to a negative translation. We instead use ideas inspired from Aschieri and Zorzi \cite{AschieriZorzi}, where it is provided a new technique to show strong normalization for G\"odel's system $\SystemTG$. The general idea is to prove not the strong normalization of the target system, but rather of a simple non-deterministic version of it, which includes in some obvious way the original system. Namely, in \cite{AschieriZorzi}, a non-deterministic version $\nSystemT$ of $\SystemTG$ is considered, which contains a non-deterministic iterator and a de' Liguoro-Piperno non-deterministic choice operator. The point is that the strong normalization of $\nSystemT$ is ``easy'' and $\SystemTG$ is obviously contained in $\nSystemT$. We shall exploit this technique of ``non-deterministic immersion'': we are not going to prove directly the strong normalization of $\HA+\EM_{1}$, but instead of a straightforward non-deterministic variation of it, the system $\HA+\NEM$, by a standard reducibility method. This latter system has exactly the same rules of the former, but for the excluded middle which is interpreted by a non-deterministic choice operator. Again, it will be ``obvious'' that the system $\HA+\EM_{1}$  is ``included'' in $\HA+\NEM$ and strong normalization will easily follow.

  The advantage of our normalization technique over the realizability one in \cite{ABB} is that it is considerably simpler; however, it is not just a simplification of the latter, but an essentially different proof.  Actually, it  shows the strong normalization of $\HA+\NEM$, which contains much more reductions than those in $\HA+\EM_{1}$, actually all the logically consistent ones which are not in $\HA+\EM_{1}$ and even some without direct logical meaning. Moreover, it is clear the realizability of \cite{ABB} needs major new ideas to be extended to $\mathsf{PA}$, while our technique seems more promising as a  first tool to prove the strong normalization of a future extension of our Curry-Howard correspondence for $\HA+\EM_{1}$ to full $\mathsf{PA}$.  Last, our method has the familiar form of a reducibility proof for an \emph{intuitionistic} system, which we find quite surprising: we are in a \emph{classical} setting, after all!

\subsection{Plan of the Paper}

This is the plan of the paper. In \S \ref{section-system} we introduce a type theoretical version of intuitionistic arithmetic $\HA$ extended with $\EM_{1}$. In \S \ref{section-systemNEM}, we present the system $\HA+\NEM$ and prove that it can simulate the reduction rules of $\HA+\EM_{1}$.
In \S \ref{section-reducibility} we define reducibility for $\HA+\NEM$ and in \S \ref{section-reducibilityproperties} we prove its main properties. Finally, in \S  \ref{section-adequacy} we prove that this reducibility is sound for $\HA+\EM_{1}$. As a corollary, we deduce that $\HA+\EM_{1}$ is strongly normalizing.

\section{The System $\HA+\EM_{1}$}
\label{section-system}
In this section we formalize intuitionistic Arithmetic $\HA$, and we add an operator $\E{a}{}{}$ formalizing $\EM_{1}$. We start with the language of formulas.

\begin{definition}[Language of $\HA + \EM_1$]\label{definition-languagear}
The language $\Language$ of $\HA + \EM_1$ is defined as follows.
\begin{enumerate}

\item
The terms of $\Language$ are inductively defined as either variables $\alpha, \beta,\ldots$ or $0$ or $\suc(t)$ with $t\in\Language$. A numeral is a term of the form $\suc\ldots \suc 0$. \\

\item
There is one symbol $\mathcal{P}$ for every primitive recursive relation over $\mathbb{N}$. The atomic formulas of $\Language$ are all the expressions of the form $\mathcal{P}(t_{1}, \ldots, t_{n})$ such that $t_{1}, \ldots, t_{n}$ are terms of $\Language$ and $n$ is the arity of $\mathcal{P}$. Atomic formulas will also be denoted as $\emp{}, \mathsf{Q}, \emp{i}, \ldots$. \\

\item
The formulas of $\Language$ are built from atomic formulas of $\Language$ by the connectives $\lor,\land,\rightarrow \forall,\exists$ as usual, with quantifiers ranging over numeric variables $\alpha^{\Nat}, \beta^{\Nat}, \ldots$.\\

\end{enumerate}

\end{definition}

\begin{figure*}[!htb]
\footnotesize{


\begin{description}

\item[Grammar of Untyped Terms]
\[t,u, v::=\ x\  |\ tu\ |\ tm\ |\ \lambda x u\  |\ \lambda \alpha u\ |\ \langle t, u\rangle\ |\ \pi_0u\ |\ \pi_{1} u\ |\ \inj_{0}(u)\ |\ \inj_{1}(u)\  |\ t[x.u, y.v]\ |\ (m,t)\ |\ t[(\alpha, x). u]\]
\[|\ \E{a}{u}{v}\ |\ [a]\Hyp{P}{\alpha}\ |\ [a]\Wit{P}{\alpha}\ |\ \True \ |\ \rec u v m \ |\ \mathsf{r}t_{1}\ldots t_{n}\]
where $m$ ranges over terms of $\Language$, $x$ over proof terms variables and $a$ over hypothesis variables and $\mathsf{r}$ is a constant.
\comment{We also assume that the term formation rules are applied in such a way that in each term $t$, if $t$ contains $[a]\Wit{P}{\alpha}$ or $[a]\Hyp{P}{\alpha}$ and $t$ contains $[a]\Wit{Q}{\alpha}$ or $[a]\Wit{Q}{\alpha}$,  then $\mathsf{P}=\mathsf{Q}$.}
We also assume that in the term $\E{a}{u}{v}$, there is some predicate $\emp{}$, such that $a$ occurs free in $u$ only in subterms of the form $[a]\Hyp{\emp{}}{\alpha}$ and $a$ occurs free in $v$ only in subterms of the form $[a]\Wit{\emp{}}{\alpha}$.
\item[Contexts] With $\Gamma$ we denote contexts of the form $e_1:A_1, \ldots, e_n:A_n$, where each $e_{i}$ is either a proof-term variable $x, y, z\ldots$ or a $\EM_{1}$ hypothesis variable $a, b, \ldots$, and $e_{i}\neq e_{j}$ for $i\neq j$.

\comment{or $\wit \beta$ (for some individual variable $\beta$) and $A_1,\ldots, A_n$ formulas of $\Language$.}
\item[Axioms] 
$\begin{array}{c}   \Gamma, x:{A}\vdash x: A
\end{array}\ \ \ \ $
$\begin{array}{c}   \Gamma, a:{\forall \alpha^{\Nat} \emp{}}\vdash  [a] \Hyp{\emp{}}{\alpha}:  \forall\alpha^{\Nat} \emp{}
\end{array}\ \ \ \ $
$\begin{array}{c}   \Gamma, a:{\exists \alpha^{\Nat} \lnot\emp{}}\vdash [a]\Wit{\emp}{\alpha}:  \exists\alpha^{\Nat} \lnot \emp{}
\end{array}$\\

\item[Conjunction] 
$\begin{array}{c}  \Gamma \vdash u:  A\ \ \ \Gamma\vdash t: B\\ \hline \Gamma\vdash \langle
u,t\rangle:
A\wedge B
\end{array}\ \ \ \ $
$\begin{array}{c} \Gamma \vdash u: A\wedge B\\ \hline\Gamma \vdash\pi_0u: A
\end{array}\ \ \ \ $
$\begin{array}{c}  \Gamma \vdash u: A\wedge B\\ \hline \Gamma\vdash\pi_1 u: B
\end{array}$\\\\

\item[Implication] 
$\begin{array}{c}  \Gamma\vdash t: A\rightarrow B\ \ \ \Gamma\vdash u:A \\ \hline
\Gamma\vdash tu:B
\end{array}\ \ \ \ $
$\begin{array}{c}  \Gamma, x:A \vdash u: B\\ \hline \Gamma\vdash \lambda x u:
A\rightarrow B
\end{array}$\\\\
\item[Disjunction Intro.] 
$\begin{array}{c}  \Gamma \vdash u: A\\ \hline \Gamma\vdash \inj_{0}(u): A\vee B
\end{array}\ \ \ \ $
$\begin{array}{c}  \Gamma \vdash u: B\\ \hline \Gamma\vdash\inj_{1}(u): A\vee B
\end{array}$\\\\

\item[Disjunction Elimination] $\begin{array}{c} \Gamma\vdash u: A\vee B\ \ \ \Gamma, x: A \vdash w_1: C\ \ \ \Gamma, x:B\vdash w_2:
C\\ \hline \Gamma\vdash  u [x.w_{1}, x.w_{2}]: C
\end{array}$\\\\

\item[Universal Quantification] 
$\begin{array}{c} \Gamma \vdash u:\forall \alpha^{\Nat} A\\ \hline  \Gamma\vdash um: A[m/\alpha]
\end{array} $
$\begin{array}{c}  \Gamma \vdash u: A\\ \hline \Gamma\vdash \lambda \alpha u:
\forall \alpha^{\Nat} A
\end{array}$\\

where $m$ is a term of  the language $\Language$ and $\alpha$ does not occur
free in any formula $B$ occurring in $\Gamma$.\\

\item[Existential Quantification] 
$\begin{array}{c}\Gamma\vdash  u: A[m/\alpha]\\ \hline \Gamma\vdash (
m,u):
\exists
\alpha^\Nat. A
\end{array}$ \ \ \ \
$\begin{array}{c} \Gamma\vdash u: \exists \alpha^\Nat A\ \ \ \Gamma, x: A \vdash t:C\\
\hline
\Gamma\vdash u [(\alpha, x). t]: C
\end{array} $\\

where $\alpha$ is not free in $C$
nor in any formula $B$ occurring in $\Gamma$.\\

\item[Induction] 
$\begin{array}{c} \Gamma\vdash u: A(0)\ \ \ \Gamma\vdash v:\forall \alpha^{\Nat}.\,
A(\alpha)\rightarrow A(\suc(\alpha))\\ \hline \Gamma\vdash \rec u v t :
 A(t)
\end{array}\ \ \ \ $\\ \\
where $t$ is any term of the language $\Language$.
\\

\item[Post Rules] 
$\begin{array}{c}  \Gamma\vdash u_1: \emp{1}\ \Gamma\vdash u_2: \emp{2}\ \cdots \ \Gamma\vdash u_n:
\emp{n}\\ \hline\Gamma\vdash u: \emp{}
\end{array}$

where $\emp{1},\emp{2},\ldots,\emp{n}, \emp{}$ are atomic
formulas  and the rule is a Post rule for equality, for a Peano axiom or a primitive recursive relation and if $n>0$, $u=\mathsf{r} u_{1}\ldots u_{n}$, otherwise $u=\True$.  \\

\item[$\EM_{1}$]$\begin{array}{c} \Gamma, a: \forall \alpha^{\Nat} \emp{} \vdash w_1: C\qquad \ \ \ \ \ \Gamma, a: \exists \alpha^{\Nat}\lnot \emp{} \vdash w_2:
C\\ \hline \Gamma\vdash  \E{a}{w_{1}}{w_{2}} : C
\end{array}$\\\\

\end{description}
}


\caption{Term Assignment Rules for $\HA+\EM_{1}$}\label{fig:D}
\end{figure*}

$\HA+\EM_{1}$  is formally described in figure \ref{fig:D}. It is a standard natural deduction system (see \cite{Sorensen}, for example), with introduction and elimination rule for each connective and induction rules for integers, together with a term assignment in the spirit of Curry-Howard correspondence. For detailed descriptions and explanations, we refer to  \cite{ABB,Birolo}.

 \comment{The EM1-rule $\EM{a}{u}{v} : C$ takes the form of an elimination rule for $\vee$ applied to the axiom $\forall \alpha. \mathsf{P} \vee \exists \alpha. \lnot \mathsf{P}$: however, we do not explicitly write the axiom $\forall \alpha. \mathsf{P} \vee \exists \alpha. \lnot \mathsf{P}$, and we only write the proofs $u$, $v$ of $C$ from $\forall \alpha. \mathsf{P}$ and from $\exists \alpha. \lnot \mathsf{P}$. These two assumptions are both associated to $a$, in the way explained below.}
 
We replace purely universal axioms (i.e., $\Pi^{0}_{1}$-axioms) with Post rules (as in Prawitz \cite{Prawitz}), which are inferences of the form
$$\begin{array}{c}  \Gamma\vdash \emp{1}\ \ \Gamma\vdash \emp{2}\ \cdots \ \ \Gamma\vdash
\emp{n}\\ \hline\Gamma\vdash  \emp{}
\end{array}$$
where $\emp{1}, \ldots, \emp{n}, \emp{}$ are atomic formulas of $\Language$  such that for every substitution $\sigma=[t_{1}/\alpha_{1}, \ldots, t_{k}/\alpha_{k}]$ of closed terms $t_{1}, \ldots, t_{k}$ of $\Language$,  if $\emp{1}\sigma=\ldots=\emp{n}\sigma=\True$ then $\emp{}\sigma=\True$. Let now $\eq$ be the symbol for the binary relation of equality between natural numbers. Among the Post rules, we have the  Peano axioms
$$\begin{array}{c}  \Gamma \vdash\eq(\suc t_1,\suc t_2)\\ \hline\Gamma\vdash  \eq(t_1,t_2)
\end{array}\qquad
\begin{array}{c}   \Gamma\vdash  \eq (0, \suc t)\\ \hline \Gamma \vdash \bot
\end{array}$$
and axioms of equality
$$
\begin{array}{c}  \\ \hline\Gamma\vdash   \eq (t, t)
\end{array}\qquad
\begin{array}{c} \Gamma\vdash \eq(t_1, t_2)\ \ \Gamma\vdash \eq(t_2, t_{3})  \\ \hline\Gamma\vdash   \eq(t_1, t_{3})
\end{array}\qquad
\begin{array}{c} \Gamma \vdash \emp{}[t_1/\alpha]\ \ \Gamma \vdash \eq(t_1, t_2)  \\ \hline\Gamma\vdash  \emp{}[t_{2}/\alpha]
\end{array}
$$
We  also have a Post rule for the defining axioms of each primitive recursive relation, for example the false $0$-ary relation $\bot$, addition, multiplication:
$$
\begin{array}{c} \Gamma\vdash\bot  \\ \hline\Gamma\vdash   \emp{}
\end{array}\qquad
\begin{array}{c}   \\ \hline\Gamma\vdash   \add(t, 0, t)
\end{array}\qquad
\begin{array}{c} \Gamma \vdash \add(t_{1}, t_{2}, t_{3} )  \\ \hline\Gamma\vdash  \add(t_{1}, \suc t_{2}, \suc t_{3})
\end{array}$$
$$
\begin{array}{c}   \\ \hline\Gamma\vdash   \mult(t, 0, 0)
\end{array}\qquad
\begin{array}{c} \Gamma \vdash \mult(t_{1}, t_{2}, t_{3} )\ \ \Gamma\vdash \add(t_{3}, t_{1}, t_{4})  \\ \hline\Gamma\vdash  \mult(t_{1}, \suc t_{2}, t_{4})
\end{array}
$$

We assume that in the proof terms three distinct classes of variables appear: one for proof terms, denoted usually as $x, y,\ldots$; one for quantified variables of the formula language $\Language$ of $\HA+\EM_{1}$, denoted usually as $\alpha, \beta, \ldots$; one for the pair of hypotheses bound by $\EM_{1}$, denoted usually as $a, b, \ldots$.  With $\lnot \mathsf{P}$ we denote the atomic predicate equivalent to the boolean negation of $\mathsf{P}$. In the term $\E{a}{u}{v}$, any free occurrence of $a$ in $u$ occurs in an expression $[a]\Hyp{P}{\alpha}$, and denotes an assumption $\forall \alpha^{\Nat} \mathsf{P}$. Any free occurrence of $a$ in $v$ occurs in an expression $[a]\Wit{P}{\alpha}$, and denotes an assumption $\exists \alpha^{\Nat} \lnot \mathsf{P}$. All the occurrences of $a$ in $u$ and $v$ are bound, and we assume the usual renaming rules and alpha equivalences to avoid capture of variables.  In the terms $[a]\Hyp{P}{\alpha}$ and $[a]\Wit{P}{\alpha}$ the free variables are $a$ and those of $\emp{}$ minus $\alpha$.

If $\Gamma \vdash t: A$, $t$ is said to be a typed proof term; if  $t$ contains as free variables only $\EM_{1}$-hypothesis variables $a_{1}, \ldots, a_{n}$ and each occurrence of them is of the form $[a_{i}]\Hyp{\emp{i}}{\alpha}$, for some $i$ and $\emp{i}$, then $t$ is said to be \emph{quasi-closed}.  $\sn$ is the set of strongly normalizable untyped proof terms and $\nf$ is the set of normal untyped proof terms, with respect to the reduction relation $\mapsto$ given in figure \ref{fig:F}.

We are now going to explain the reduction rules for the proof terms of $\HA+\EM_{1}$ (with $\mapsto^{*}$ we shall denote the reflexive and transitive closure of the one-step reduction $\mapsto$). We find among them the ordinary reductions of intuitionistic Arithmetic for the logical connectives and induction. Permutation rules for $\EM_{1}$ are an instance of Prawitz's permutation rules for $\vee$-elimination; the full permutations rules for $\lor$ and $\exists$ eliminations are not needed, since $\HA+\EM_{1}$ cannot have the subformula property, which is one the main logical reasons why permutations are usually considered. Given these rules and those for $\EM_{1}$, it is possible to prove syntactically that a normal form proof of a $\Sigma_{0}^{1}$-formula is constructive, i.e. it always ends with an $\exists$-introduction \cite{Birolo}.   Raising an exception $n$ in $\E{a}{u}{v}$ removes all occurrences of assumptions $[a]\Wit{P}{\alpha}$ in $v$. We define first an operation removing.

\begin{definition}Suppose $v$ is any term. We define $v[a:=n]$
as the term obtained from $v$ by replacing each  subterm $[a]\Wit{P}{\alpha}$ corresponding to a free occurrence of $a$ in $v$ by $(n, \True)$. \comment{ if $\emp{}[n/\alpha]=\False$, and by $(n,[a]\Hyp{\,\alpha=0}{\alpha}\suc 0)$, otherwise.} \end{definition}

\comment{
 \begin{definition}Suppose there is a closed formula $\exists \alpha^{\Nat}\emp{}$ such that $a$ occurs free in $v$ only in subterms of the form $[a]\Wit{\emp{}}{\alpha}$ and let $n$ be a numeral; we set
 $$v[a:=n]\overset{def}{=}v[(n,\True)/[a]\Wit{\emp{}}{\alpha}]$$
 if $\emp{}[n/\alpha]=\False$; otherwise, we set
 $$v[a:=n]\overset{def}{=}v[(n,[a']\Hyp{\,\alpha=0}{\alpha}\suc 0)/[a]\Wit{\emp{}}{\alpha}]$$
 for some fresh variable $a'$.
 In other terms, $v[a:=n]$ is obtained from $v$ by replacing every subterm $[a]\Wit{P}{\alpha}$ corresponding to a free occurrence of $a$ in $v$ by either $(n, \True)$ or $(n,[a']\Hyp{\,\alpha=0}{\alpha}\suc 0)$. \end{definition}}
 
 The rules for $\EM_{1}$ translate the informal idea of learning by trial and error. The first $\EM_{1}$-reduction: $([a]\Hyp{P}{\alpha}) n \mapsto \True$ if $\emp{}[n/\alpha]=\True$, says that whenever we use an instance $\emp{}[n/\alpha]$ of the assumption $\forall \alpha^{\Nat} \mathsf{P}$, we check it, and if the instance is true we replace it with its canonical proof.  The second $\EM_{1}$-reduction: $\E{a}{u}{v}\mapsto u$, says that if, using the first reduction, we are able to remove all the instances of the assumption $[a]\Hyp{P}{\alpha} : \forall \alpha^{\Nat} \mathsf{P} $ in $u$, then the assumption is unnecessary and the proof $\E{a}{u}{v}$ may be simplified to $u$. In this case the exceptional part $v$ of $\E{a}{u}{v}$ is never used. The third $\EM_{1}$-reduction: $\E{a}{u}{v}\mapsto v[a:=n]$, if $[a]\Hyp{\emp{}}{\alpha} n$ occurs in $u$ and $\emp{}[n/\alpha]=\False$, says that if we check an instance $[a]\Hyp{\emp{}}{\alpha} n : \mathsf{P}[n/\alpha]$ of the assumption $\forall \alpha^{\Nat} \mathsf{P} $, and we find that the assumption is wrong, then we raise the exception $n$ and we start the exceptional part $v$ of $\E{a}{u}{v}$. Raising an exception is a non-deterministic operation (we may have two or more exception to choose) and has no effect outside $\E{a}{u}{v}$. 
 
 As pointed out to us by H. Herbelin, the whole term $\E{a}{u}{v}$ can also be expressed in a standard way by the constructs $\mathsf{raise}$ and $\mathsf{try}\ldots \mathsf{with}\ldots$ in the $\mathsf{CAML}$ programming language and $[a]\Hyp{P}{\alpha}$ corresponds to a declaration of an exception $a$ which is raised in the left branch of $\mathsf{try}\ldots \mathsf{with}\ldots$ when applied to an integer falsifying the hypothesis $\forall \alpha^{\Nat} \emp{}$.   Also de Groote pointed out in \cite{deGrooteex} the analogy between this kind of reductions and the exception handling mechanism in the Standard $\mathsf{ML}$. Our reductions for $\EM_{1}$, however, differ substantially from de Groote's ones in the way exceptions are menaged and raised. We use a ``truth-based'' mechanism of exception raising, while de Groote's uses a ``proof-based'' one. That is, we raise an exception only when we have falsified an hypothesis, while de Groote raises an exception when a proof of the contrary of the hypothesis is found (and given the presence of possibly many other hypotheses this does not imply that the hypothesis is actually false). Therefore, our reductions belong to the same family of Interactive realizability \cite{ABF,AschieriCSL}, while de Groote's belong to the same family of the Griffin, Krivine and G\"odel double-negation approach to classical proofs. 

\begin{figure*}[!htb]
\footnotesize{
\dlinea
\begin{description}

\item[Reduction Rules for $\HA$]
\[(\lambda x. u)t\mapsto u[t/x]\qquad (\lambda \alpha. u)m\mapsto u[m/\alpha]\]
 \[ \pi_{i}\pair{u_0}{u_1}\mapsto u_i, \mbox{ for i=0,1}\]
\[\inj_{i}(u)[x_{1}.t_{1}, x_{2}.t_{2}]\mapsto t_{i}[u/x_{i}], \mbox{ for i=0,1} \]
\[(n, u)[(\alpha,x).v]\mapsto v[n/\alpha][u/x], \mbox{ for each numeral $n$} \]
\[\rec u v 0 \mapsto u\]
\[\rec u v (\suc n) \mapsto v n (\rec u v n), \mbox{ for each numeral $n$} \]

\item[Permutation Rules for $\EM_{1}$]
\[(\E{a}{u}{v}) w \mapsto \E{a}{uw}{vw} \]
\[\pi_{i}(\E{a}{u}{v})  \mapsto \E{a}{\pi_{i}u}{\pi_{i}v} \]
\[(\E{a}{u}{v})[x.w_{1}, y.w_{2}] \mapsto \E{a}{u[x.w_{1}, y.w_{2}]}{v[x.w_{1}, y.w_{2}]} \]
\[(\E{a}{u}{v})[(\alpha, x).w] \mapsto \E{a}{u[(\alpha, x).w]}{v[(\alpha, x).w]} \]
\item[Reduction Rules for $\EM_{1}$]
\[([a]\Hyp{P}{\alpha}) n \mapsto \True, \mbox{ if } \emp{}[n/\alpha]=\True\]
\[\E{a}{u}{v}\mapsto u,\ \mbox{ if $a$ does not occur free in $u$ }\]
\[\E{a}{u}{v}\mapsto v[a:=n]\comment{\overset{def}{=}v[(n,\True)/[a]\Wit{\emp{}}{\alpha}]},\ \mbox{ if $[a]\Hyp{\emp{}}{\alpha} n$ occurs in $u$ and $\emp{}[n/\alpha]=\False$}\]
\comment{\item[Extra Reduction Rules]
\[[a]\Wit{P}{\alpha}  \mapsto (n, \True) \mbox{ for each numeral $n$ }\]
\[\E{a} u v\mapsto u\]
\[\E{a} u v\mapsto v\]}
\end{description}}
\dlinea
\caption{Reduction Rules for $\HA$ + $\EM_{1}$}\label{fig:F}
\end{figure*}

\section{The System $\HA+\NEM$}\label{section-systemNEM}

In this section we introduce the non-deterministic system $\HA+\NEM$, which is still a standard natural deduction system for Heyting Arithmetic with $\EM_{1}$. The only syntactical difference with the system $\HA+\EM_{1}$ lies in the shape of proof terms, and is really tiny: the proof terms for $\EM_{1}$ and $\EM_{1}$-hypotheses lose the hypothesis variables used to name them. Thus the grammar of untyped proof terms of $\HA+\NEM$ is defined to be  the following: 
\begin{description}

\item[Grammar of Untyped Terms of $\HA+\NEM$]
\[t,u, v::=\ x\  |\ tu\ |\ tm\ |\ \lambda x u\  |\ \lambda \alpha u\ |\ \langle t, u\rangle\ |\ \pi_0u\ |\ \pi_{1} u\ |\ \inj_{0}(u)\ |\ \inj_{1}(u)\  |\ t[x.u, y.v]\ |\ (m,t)\ |\ t[(\alpha, x). u]\]
\[|\ \E{}{u}{v}\ |\ \Hyp{P}{\alpha}\ |\ \Wit{P}{\alpha}\ |\ \True \ |\ \rec u v m \ |\ \mathsf{r}t_{1}\ldots t_{n}\]
where $m$ ranges over terms of $\Language$, $x$ over proof terms variables and $a$ over hypothesis variables.\\
\end{description}
The term assignment rules of $\HA+\NEM$ are exactly the same of $\HA+\EM_{1}$, but for the ones for $\EM_{1}$-hypotheses and $\EM_{1}$, which (obviously) become:\\

\begin{description}

\item[Axioms] 
$\begin{array}{c}   \Gamma, a:{\forall \alpha^{\Nat} \emp{}}\vdash  \Hyp{\emp{}}{\alpha}:  \forall\alpha^{\Nat} \emp{}
\end{array}\ \ \ \ $
$\begin{array}{c}   \Gamma, a:{\exists \alpha^{\Nat} \lnot\emp{}}\vdash \Wit{\emp}{\alpha}:  \exists\alpha^{\Nat} \lnot \emp{}
\end{array}$ \\

\item[$\NEM$]$\begin{array}{c} \Gamma, a: \forall \alpha^{\Nat} \emp{} \vdash w_1: C\ \ \ \Gamma, a: \exists \alpha^{\Nat}\lnot \emp{} \vdash w_2:
C\\ \hline \Gamma\vdash  \E{}{w_{1}}{w_{2}} : C\\\\\\
\end{array}$
\end{description}

The reduction rules for the terms of $\HA+\NEM$ are defined in figure \ref{fig:FN} and  are those of the first two groups for $\HA+\EM_{1}$, plus new non-deterministic rules for $\NEM$ (with $\redn^{*}$ we shall denote the reflexive and transitive closure of the one-step reduction $\redn$).  Thus, in the system $\HA+\NEM$ the operator $\E{}{}{}$ behaves as a standard de' Liguoro-Piperno non-deterministic choice operator (see \cite{DeLPip1,MarghiUgo}). The term $\Wit{P}{\alpha}$ behaves as a ``search'' operator, which spans non-deterministically all natural numbers as possible witnesses of $\exists \alpha^{\Nat}\lnot\emp{}$. The reduction tree of a strongly normalizable term with respect to $\redn$ is no more finite, but still well-founded. It is well-known that it is possible to assign to each node of a well-founded tree an ordinal number, in such a way it decreases passing from a node to any of its sons. We will call the \emph{ordinal size} of a term $t\in\sn$ the ordinal number assigned to the root of its reduction tree and we denote it by $h(t)$; thus, if $t\redn u$, then $h(t)>h(u)$. To fix ideas, one may define $h(t):=\mathsf{sup}\{ h(u)+1\ |\ t\mapsto u\}$. \begin{figure*}[!htb]
\footnotesize{
\dlinea
\begin{description}
\item[Reduction Rules for $\HA$]
\[(\lambda x. u)t\redn u[t/x]\qquad (\lambda \alpha. u)t\redn u[t/\alpha]\]
 \[ \pi_{i}\pair{u_0}{u_1}\redn u_i, \mbox{ for $i=0,1$}\]
\[\inj_{i}(u)[x_{1}.t_{1}, x_{2}.t_{2}]\redn t_{i}[u/x_{i}], \mbox{ for $i=0,1$} \]
\[(n, u)[(\alpha,x).v]\redn v[n/\alpha][u/x], \mbox{ for each numeral $n$} \]
\[\rec u v 0 \redn u\]
\[\rec u v (\suc n) \redn v n (\rec u v n), \mbox{ for each numeral $n$} \]

\item[Permutation Rules for $\NEM$]
\[(\E{}{u}{v}) w \redn \E{}{uw}{vw} \]
\[\pi_{i}(\E{}{u}{v})  \redn \E{}{\pi_{i}u}{\pi_{i}v} \]
\[(\E{}{u}{v})[x.w_{1}, y.w_{2}] \redn \E{}{u[x.w_{1}, y.w_{2}]}{v[x.w_{1}, y.w_{2}]} \]
\[(\E{}{u}{v})[(\alpha, x).w] \redn \E{}{u[(\alpha, x).w]}{v[(\alpha, x).w]} \]
\item[Reduction Rules for $\NEM$]
\[(\Hyp{P}{\alpha}) n \redn \True, \mbox{ if } \emp{}[n/\alpha]=\True\]
\[\Wit{P}{\alpha}\redn (n, \True), \mbox{ for every numeral $n$}\]
\[\E{}{u}{v}\redn u \]
\[\E{}{u}{v}\redn v\]

\end{description}}
\dlinea
\caption{Reduction Rules for $\HA$ + $\NEM$}\label{fig:FN}
\end{figure*}

We now define the obvious translation mapping untyped proof terms of $\HA+\EM_{1}$ into untyped terms of $\HA+\NEM$, which just erases every occurrence of every $\EM_{1}$-hypothesis variable $a$. 

\begin{definition}[Translation of untyped proof terms of $\HA+\EM_{1}$ into $\HA+\NEM$]\mbox{}
We define a translation $\trans{\_}$ mapping untyped proof terms of $\HA+\EM_{1}$ into untyped proof terms of $\HA+\NEM$: $\trans{t}$ is defined as the term of $\HA+\NEM$ obtained from $t$ by erasing every expression of the form $[a]$ and replacing each occurrence of the symbol $\E{a}{}{}$ with $\E{}{}{}$.
\end{definition}

We now show that the reduction relation $\redn$ for the proof terms of $\HA+\NEM$ can easily simulate the reduction relation $\mapsto$ for the terms of $\HA+\EM_{1}$. This is trivial for the proper reductions of $\HA$ and the permutative reductions for $\EM_{1}$, while the reduction rules for the terms of the form $\E{a}{u}{v}$  can be plainly simulated by $\redn$  with  non-deterministic guesses. In particular, each reduction step between  terms of $\HA+\EM_{1}$ corresponds to \emph{at least} a step between their translations:

\begin{proposition}[Preservation of the Reduction Relation $\mapsto$ by $\redn$]\label{proposition-preservation}
Let $v$ be any untyped proof term of $\HA+\EM_{1}$. Then $v\mapsto w\implies \trans{v}\redn^{+} \trans{w}$
\end{proposition}
\textit{Proof}. It is sufficient to prove the proposition when $v$ is a redex $r$. We have several possibilities, almost all trivial, and we choose only some representative cases:
\begin{enumerate}
\item $r=(\lambda x\, u)t\mapsto u[t/x]$. 
We verify indeed that $$ \trans{((\lambda x\, u)t)}=(\lambda x\, \trans{u})\trans{t}\redn \trans{u}[\trans{t}/x]= \trans{u[t/x]}$$

\comment{\item $r=\pair{u_{0}}{u_{1}}\pi_{i}\redn u_{i}$. 
We verify indeed that 
$$\trans{(\pair{u_{0}}{u_{1}}\pi_{i})}=\pair{\trans{u}_{0}}{\trans{u}_{1}}\pi_{i}\redn \trans{u}_{i}$$}

\item $r=(\E{a}{u}{v})w\mapsto \E{a}{uw}{vw}$. 
\noindent
We verify indeed  that 
$$\trans{((\E{a}{u}{v})w)}=(\E{}{\trans{u}}{\trans{v}})\trans{w}\redn \E{}{\trans{u}\trans{w}}{\trans{v}\trans{w}}\redn \trans{(\E{a}{uw}{vw})}$$
\item $r=\E{a}{u}{v}\mapsto v[a:=n]$. We verify indeed -- by choosing the appropriate reduction rule for $\E{}{}{}$ and applying repeatedly the reduction rule $\Wit{P}{\alpha}\redn (n, \True)$ -- that 
$$\trans{(\E{a}{u}{v})}= \E{}{\trans{u}}{\trans{v}}\redn \trans{v}\redn^{*} \trans{(v[a:=n])} $$
\end{enumerate}

\section{Reducibility}\label{section-reducibility}

We now want to prove the strong normalization theorem for $\HA+\NEM$: every term $t$ which is typed in $\HA+\NEM$ is strongly normalizable. We use a simple extension of the reducibility method of Tait-Girard \cite{Girard}.

\begin{definition}[Reducibility]
\label{definition-reducibility}
Assume $t$ is a term in the grammar of untyped terms of $\HA+\NEM$ and $C$ is a formula of $\Language$. We define the relation $t\red C$ (``$t$ is reducible of type $C$'') by induction and by cases according to the form of $C$:

\begin{enumerate}
\item
$t\red \emp{}$ if and only if 
$t\in\sn$\\

\item
$t\red {A\wedge B}$ if and only if $\pi_0t \red {A}$ and $\pi_1t\red {B}$\\

\item
$t\red {A\rightarrow B}$ if and only if for all $u$, if $u\red {A}$,
then $tu\red {B}$\\

\item
$t\red {A\vee B}$  if and only if $t\in\sn$ and $t\redn^{*} \inj_{0}(u)$ implies $u\red A$ and $t\redn^{*} \inj_{1}(u)$ implies $u\red B$\\

\item
$t\red {\forall \alpha^{\Nat} A}$ if and only if for every term $n$ of $\Language$,
$t{n}\red A[{n}/\alpha]$\\
\item

$t\red \exists \alpha^{\Nat} A$ if and only if $t\in\sn$ and for every term $n$ of $\Language$, if $t\redn^{*} (n,u)$, then $u \red A[{n}/\alpha]$\\
\end{enumerate}
\end{definition}

\section{Properties of Reducible Terms}\label{section-reducibilityproperties}

In this section we prove that the set of reducible terms for a given formula $C$ satisfies the usual properties of a Girard's reducibility candidate. 

Following \cite{Girard}, neutral terms are terms that are not ``values''  and need to be further computed.

\begin{definition}[Neutrality]
A proof term  is neutral if it is not of the form $\lambda x\, u$ or $\lambda \alpha\, u$ or $\pair{u}{t}$ or $\inj_{i}(u)$ or $(t, u)$ or $\E{}{u}{v}$ or $\Hyp{P}{\alpha}$.
\end{definition}

\begin{definition}[Reducibility Candidates] Extending the approach of \cite{Girard}, we define four properties \cruno,\ \crdue,  \crtre, \crquattro\ of reducible terms $t$:\\

\cruno\ If $t\red A$, then $t\in \sn$.\\

\crdue\ If $t \red A$ and $t\redn^{*} t'$, then $t' \red A$.\\

\crtre\  If $t$ is neutral and  for every $t'$, $t\redn t'$ implies $t'\red A$, then $t \red A$.\\

\crquattro\ $t=\E{}{u}{v}\red A$ if and only if $u\red A$ and $v\red A$.\\
\end{definition}

We now prove, as usual, that every term $t$ possesses the reducibility candidate properties. The arguments for establishing \cruno, \crdue, \crtre, are standard (see 	\cite{Girard}).

\begin{proposition}
Let $t$ be a term of $\HA+\NEM$. Then $t$ has the properties \cruno, \crdue, \crtre, \crquattro.
\end{proposition}
\textit{Proof}. By induction on $C$. 
\begin{itemize}
\item $C$ is atomic. Then $t\red C$ means $t\in\sn$. Therefore the thesis is trivial. \\

\item $C=A\rightarrow B$.\\

\cruno. Suppose $t\red A\rightarrow B$. By induction hypothesis \crtre,  for any variable $x$, we have $x\red A$. Therefore, $tx\red B$, and by \cruno, $tx\in\sn$, and thus $t\in\sn$.\\
 
 \crdue. Suppose $t\red A\rightarrow B$ and $t\redn t'$. Let $u\red A$: we have to show $t'u\red B$. Since $tu\red B$ and $tu\redn t'u$, we have by the induction hypothesis \crdue\ that $t'u\red B$.  \\
 
 \crtre.\  Assume $t$ is neutral and $t\redn t'$ implies $t'\red A\rightarrow B$. Suppose $u\red A$; we have to show that $tu \red B$. We proceed by induction on the ordinal height of the reduction tree of $u$ ($u\in\sn$ by induction hypothesis \cruno).  By induction hypothesis, \crtre\ holds for the type $B$. So assume $tu \redn z$; it is enough to show that $z\red B$.  If $z=t'u$, with $t \redn t'$, then by hypothesis $t'\red A\rightarrow B$, so $z\red B$. If $z=tu'$, with $u\redn u'$, by induction hypothesis \crdue\ $u'\red A$, and therefore $z\red B$ by the induction hypothesis relative to the size of the reduction tree of $u'$. There are no other cases since $t$ is neutral.\\
 
 \crquattro.\ $\Rightarrow$). Suppose $t=\E{}{u}{v}\red A\rightarrow B$. Since $t\redn u$, $t\redn v$, by \crdue, $u\red A\rightarrow B$ and $v\red A\rightarrow B$.\\
 $\Leftarrow$). Suppose $u\red A\rightarrow B$ and $v\red A\rightarrow B$. Let $t\red A$. We show by triple induction on the ordinal heights of the reduction trees of $u, v, t$ (they are all  in $\sn$ by \cruno) that $(\E{}{u}{v})t\red B$. By induction hypothesis \crtre,\ it is enough to assume $(\E{}{u}{v})t\redn z$ and show $z\red B$. If $z=ut$ or $vt$, we are done. If $z=(\E{}{u'}{v})t$ or $z=(\E{}{u}{v'})t$ or $(\E{}{u}{v})t'$, with $u\redn u'$, $v\redn v'$ and $t\redn t'$, we obtain $z\red B$ by \crdue\ and induction hypothesis. If $z=(\E{}{ut}{vt})$, by induction hypothesis \crquattro,\ $z\red B$.  \\

 \item $C=\forall \alpha^{\Nat} A$ or $C=A\land B$. Similar to the case $C=A\rightarrow B$.\\
 \item $C=A_{0}\lor A_{1}$. \\
 
 \cruno\ is trivial. \\
 
 \crdue.\ Suppose $t\red A_{0}\lor A_{1}$ and $t\redn^{*} t'$. Then $t'\in\sn$, since $t\in\sn$. Moreover, suppose $t'\redn^{*} \inj_{i}(u)$. Then also $t\redn^{*} \inj_{i}(u)$, so $u\red A_{i}$. \\
 
 \crtre.\  Assume $t$ is neutral and $t\redn t'$ implies $t'\red A_{0}\lor A_{1}$. Since $t\redn t'$ implies $t'\in\sn$, we have $t\in\sn$. Moreover, if $t\redn^{*} \inj_{i}(u)$, then, since $t$ is neutral, $t\redn t'\redn^{*}\inj_{i}(u)$ and thus $u\red A_{i}$.  \\
 
\crquattro. $\Rightarrow$). Suppose $t=\E{}{u}{v}\red A_{0}\lor A_{1}$. Since $t\redn u$, $t\redn v$, by \crdue, $u\red A_{0}\lor A_{1}$ and $v\red A_{0}\lor A_{1}$.\\
 $\Leftarrow$). Suppose $u\red A_{0}\lor A_{1}$ and $v\red A_{0}\lor A_{1}$. By \cruno, $u, v\in\sn$; therefore, $\E{}{u}{v}\in\sn$. Moreover, suppose $\E{}{u}{v}\redn^{*}\inj_{i}(w)$. Then, either $u\redn^{*}\inj_{i}(w)$ or $v\redn^{*}\inj_{i}(w)$. Thus, $w\red A_{i}$. We conclude $\E{}{u}{v}\red A_{0}\lor A_{1}$. \\
 
 \item $C=\exists \alpha^{\Nat} A$. Similar to the case $t=A_{0}\lor A_{1}$. 
 
   \end{itemize}

The next task is to prove that all introduction and elimination rules of $\HA+\NEM$ define a reducible term from a list of reducible terms for all premises (Adequacy Theorem \ref{Adequacy Theorem}). In some case that is true by definition of reducibility; we list below some non-trivial but standard cases we have to prove.

\begin{proposition}\label{proposition-somecases}\mbox{}
\begin{enumerate}
\item If for every $t\red A$, $u[t/x]\red B$, then  $\lambda x\, u\red A\rightarrow B$.
\item If for every term $m$ of $\Language$, $u[m/\alpha]\red B[m/\alpha]$, then $\lambda \alpha\, u\red \forall \alpha^{\Nat} B$.
\item If $u\red A_{0}$ and $v\red A_{1}$, then $\pi_{i}\pair{u}{v}\red  A_{i}$.
\item If $t\red A_{0}\lor A_{1}$ and for every $t_{i}\red A_{i}$ it holds $u_{i}[t_{i}/x_{i}]\red C$, then $t[x_{0}.u_{0}, x_{1}.u_{1}]\red C$.
\item If $t\red \exists \alpha^{\Nat} A$ and for every term $n$ of $\Language$ and $v\red A[n/\alpha]$ it holds $u[n/\alpha][v/x]\red C$, then $t[(\alpha, x).u]\red C$.
\end{enumerate}
\end{proposition}
\textit{Proof}.\mbox{}
\begin{enumerate}
\item As in \cite{Girard}.\\
\item As 1.\\
\item As in \cite{Girard}.\\
\item Suppose $t\red A_{0}\lor A_{1}$ and for every $t_{i}\red A_{i}$ it holds $u_{i}[t_{i}/x_{i}]\red C$. We observe that by \crtre,\ $x_{i}\red A_{i}$, and so we have $u_{i}\red A_{i}$.  Thus, in order to prove $t[x_{0}.u_{0}, x_{1}.u_{1}]\red C$, by \cruno, we can reason by triple induction on the ordinal sizes of the reduction trees of $t, u_{0}, u_{1}$. By \crtre, it suffices to show that $t[x_{0}.u_{0}, x_{1}.u_{1}]\redn z$ implies $z\red C$. If $z=t'[x_{0}.u_{0}, x_{1}.u_{1}]$ or $z=t[x_{0}.u_{0}', x_{1}.u_{1}]$ or $z=t[x_{0}.u_{0}, x_{1}.u_{1}']$, with $t\redn t'$ and $u_{i}\redn u_{i}'$, then by \crdue\ and by induction hypothesis $z\red C$. If $t=\inj_{i}(t_{i})$ and $z=u_{i}[t_{i}/x_{i}]$, then $t_{i}\red A_{i}$; therefore, $z\red C$. If $t=\E{}{w_{0}}{w_{1}}$ and 
$$z=\E{}{(w_{0}[x_{0}.u_{0}, x_{1}.u_{1}])}{(w_{1}[x_{0}.u_{0}, x_{1}.u_{1}])}$$
then, since $t=\E{}{w_{0}}{w_{1}}\redn w_{i}$, by induction hypothesis $w_{i}[x_{0}.u_{0}, x_{1}.u_{1}]\red C$ for $i=0,1$. By \crquattro, we conclude $z\red C$. 
\item Similar to 4. 
\end{enumerate}

\section{The Adequacy Theorem}\label{section-adequacy}

\begin{theorem}[Adequacy Theorem]\label{Adequacy Theorem}
Suppose that $\Gamma\vdash w: A$ in
the system $\HA + \NEM$, with
$\Gamma=x_1: {A_1},\ldots,x_n:{A_n}, \Delta$ ($\Delta$ not containing declarations of proof-term variables), and that the free variables of the formulas occurring in $\Gamma $ and $A$ are among
$\alpha_1,\ldots,\alpha_k$. For all terms $r_1,\ldots,r_k$ of $\Language$, if there are terms $t_1, \ldots, t_n$ such that
\[\text{ for  $i=1,\ldots, n$, }t_i\red A_i[{r}_1/\alpha_1\cdots
{r}_k/\alpha_k]\]
 then
\[w[t_1/x_1\cdots
t_n/x_n\  {r}_1/\alpha_1\cdots
{r}_k/\alpha_k]\red A[{r}_1/\alpha_1\cdots
{r}_k/\alpha_k]\]
\end{theorem}

\textit{Proof}.
\newcommand{\substitution} [1]         { {\overline{#1}} }

Notation: for any term $v$ and formula $B$, we denote \[v[t_1/x_1\cdots t_n/x_n\ {r}_1/\alpha_1\cdots {r}_k/\alpha_k]\] with $\substitution{v}$ and \[B[{r}_1/\alpha_1\cdots {r}_k/\alpha_k]\] with $\substitution{B}$. We proceed by induction on $w$ and cover only the case not already treated in \cite{Girard}. Consider the last rule in the derivation of $\Gamma\vdash w: A$:

\begin{enumerate}

\item If it is the rule $\Gamma \vdash \Hyp{P}{\alpha}:  \forall\alpha^{\Nat} \emp{}$, then $w=\Hyp{P}{\alpha}$ and $A= \forall\alpha^{\Nat} \emp{}$. So $\substitution{w}=\Hyp{\substitution{P}}{\alpha}$. Let $n$ be any term of $\Language$. Obviously,  $\Hyp{\substitution{P}}{\alpha}n\in \sn$; so $\Hyp{\substitution{P}}{\alpha}n \red \substitution{\emp{}}[n/\alpha]$. We conclude $\Hyp{\substitution{P}}{\alpha} \red \forall\alpha^{\Nat} \substitution{\emp{}}=\substitution{A}$.\\

\item If it is the rule $ \Gamma \vdash \Wit{P}{\alpha}:  \exists\alpha^{\Nat} \lnot \emp{}$, then $w=\Wit{P}{\alpha}$ and $A= \exists \alpha^{\Nat} \lnot \emp{}$. So $\substitution{w}=\Wit{\substitution{P}}{\alpha}$. Obviously, $\Wit{\substitution{P}}{\alpha}\in\sn$. Moreover, for every numeral $n$, we have $\Wit{\substitution{P}}{\alpha} \redn (n,\True)$ and $\True \red \lnot\substitution{\emp{}}[n/\alpha]$. We conclude $\Wit{\substitution{P}}{\alpha} \red \exists\alpha^{\Nat} \lnot\substitution{\emp{}}=\substitution{A}$.\\

\item
If it is a $\vee I$ rule, say left (the other case is symmetric), then $w=\inj_{0}(u)$, $A=B\vee C$ and $\Gamma \vdash u: B$. So, $\substitution{w}=\inj_{0}(\substitution{u})$. By induction hypothesis $\substitution{u}\red \substitution{B}$. Hence, $\substitution{u}\in\sn$. Moreover, suppose $\inj_{0}(\substitution{u}) \redn^{*} \inj_{0}(v)$. Then $\substitution{u}\redn^{*} v$ and thus by \crdue\ $v\red \substitution{B}$. We conclude $\inj_{0}(\substitution{u}) \red \substitution{B}\lor\substitution{C}= \substitution{A}$. 
\\

\item If it is a $\vee E$ rule, then
\[w=  u [x.w_1, y.w_2] \]
 and  $\Gamma \vdash u: B\vee C$, $\Gamma, x: B \vdash w_1: D$, $\Gamma, y: C \vdash w_2: D$, $A=D$.  By induction hypothesis, we have $\substitution{u}\red \substitution{B}\lor \substitution{C}$; moreover,  for every $t\red \substitution{B}$, we have $\substitution{w}_{1}[t/x]\red \substitution{B}$ and for every $t\red \substitution{C}$, we have $\substitution{w}_{2}[t/y]\red \substitution{C}$.  By proposition \ref{proposition-somecases}, we obtain $\substitution{w}=\substitution{u} [x.\substitution{w}_1, y.\substitution{w}_2]\red \substitution{C}$
\\
\item The cases $\exists I$ and $\exists E$ are similar respectively to $\lor I$ and $\lor E$.\\
\item If it is the  $\forall E$ rule, then $w=ut$, $A=B[t/\alpha]$
and $\Gamma \vdash u: \forall \alpha^{\Nat} B$. So,
$\substitution{w}=\substitution{u}\substitution{t}$.  By inductive hypothesis  $\substitution{u}\red
\forall\alpha^{\Nat} \substitution{B}$ and so $\substitution{u}\substitution{t}\red \substitution{B}[\substitution{t}/\alpha]$. \\

\item
If it is the  $\forall I$ rule, then $w=\lambda \alpha u$, $A=\forall \alpha^{\Nat} B$ and $\Gamma \vdash u: B$ (with $\alpha$ not occurring free in the formulas of $\Gamma$). So, $\substitution{w}=\lambda \alpha \substitution{u}$, since we may assume $\alpha\neq \alpha_1, \ldots, \alpha_k$. Let $t$ be a term of $\Language$; by proposition \ref{proposition-somecases}), it is enough to prove that $\substitution{u}[t/\alpha]\red \substitution{B}[{t}/\alpha]$, which amounts to show that the induction hypothesis can be applied to $u$. For this purpose, we observe that, since $\alpha\neq \alpha_1, \ldots, \alpha_k$, for $i=1, \ldots, n$ we have
\[t_i\red \substitution{A}_i=\substitution{A}_i[t/\alpha]\]

\comment{
  \item If it is the  $\exists E$ rule, then \[w=(\lambda
\alpha^{\tau}\lambda x^{|B|} t)(\pi_0u)( \pi_1u)\] $\Gamma, x:B \vdash t: A$ and
$\Gamma \vdash u: \exists \alpha^{\tau}. B$. Assume ${v} = \pi_0
\substitution{u}[s]$. Then
\[\substitution{t}[{v}/\alpha^{\tau},\pi_1\substitution{u}/
x^{|{B}|}]\red
\substitution{A}[{v}/\alpha^{\tau}]=\substitution{A}\] by
inductive hypothesis, whose application being justified by the
fact, also by induction, that $\substitution{u}\red \exists
\alpha^{\Nat}.
\substitution{B}$ and hence $\pi_1\substitution{u}\red
\substitution{B}[{v}/\alpha^{\tau}]$. We thus obtain by
$ \substitution{w} [s] =\substitution{t}[\pi_0 \substitution{u}/\alpha^{\tau}\ \pi_1\substitution{u}/x^{|B|}] [s] $ and saturation (prop. \ref{proposition-saturation}) that
\[\substitution{w} \red \substitution{A}\]

  \item If it is the $\exists I$ rule, then $w=\langle  t,u\rangle
$, $A=\exists
\alpha^{\tau}
B$, $\Gamma \vdash u: B[t/\alpha^{\tau}]$. So, $\substitution{w}=\langle
\substitution{t},\substitution{u}\rangle $; and, indeed, $\pi_1
\substitution{w}=\substitution{u}\red \substitution{B}[\pi_0\substitution{w}/\alpha^{\tau}]=\substitution{B}[\substitution{t}/\alpha^{\tau}]$ since by induction hypothesis
$\substitution{u}\red \substitution{B}[\substitution{t}/\alpha^{\tau}]$. By saturation we conclude the thesis. \\
}
  \item If it is the induction rule, then $w=
\rec u v t$, $A=B(t)$, $\Gamma \vdash u: B(0)$ and $\Gamma \vdash v:
\forall \alpha^{\Nat}. B(\alpha)\rightarrow B(\suc(\alpha))$. So,
$\substitution{w}=
\rec \substitution{u}\substitution{v}l$, for some numeral $l=\substitution{t}$.

We  prove that for all numerals $n$, $\rec \substitution{u}\substitution{v} n \red \substitution{B}({n})$. By \crtre,\ it is enough to suppose that  $\rec \substitution{u}\substitution{v} n \mapsto w$ and show that $w\red \substitution{B}({n})$.    By induction hypothesis $\substitution{u}\red
\substitution{B}(0)$ and $\substitution{v}{m}\red
\substitution{B}({m})\rightarrow
\substitution{B}({\suc(m)})$ for all terms $m$ of $\Language$.  So by \cruno,\ we can reason by triple induction on ordinal  sizes of the reduction trees of $\substitution{u}$ and $\substitution{v}$ and the size of $m$. If $n=0$ and $w=\substitution{u}$, then we are done. If $n=\suc(m)$ and $w=\substitution{v}m(\rec \substitution{u}\substitution{v}m)$, by induction hypothesis $\rec \substitution{u}\substitution{v}m\red \substitution{B}({m})$; therefore, $w\red \substitution{B}(\suc{m})$. If $w=\rec u' \substitution{v}m$, with $\substitution{u}\redn u'$, by induction hypothesis $w\red\substitution{B}(m)$. We conclude the same if  $w=\rec \substitution{u} {v}'m$, with $\substitution{v}\redn v'$. \\

\item If it is the $\NEM$ rule, then $w= \E{}{u}{v}$, $\Gamma, a: \forall \alpha^{\Nat} \emp{} \vdash u: C$ and $\Gamma, a:\exists \alpha^{\Nat}\lnot \emp{} \vdash v:
C$ and $A=C$. By induction hypothesis, $\substitution{u}, \substitution{v}\red \substitution{C}$. By  \crquattro, we conclude $\substitution{w}=\E{}{\substitution{u}}{\substitution{v}}\red \substitution{C}$. \\

  \item If it is a Post rule, the thesis follows immediately by induction hypothesis.

\end{enumerate}


As corollary, one obtain strong normalization for $\HA+\NEM$.

\begin{corollary}[Strong Normalization for $\HA+\NEM$] Suppose $\Gamma\vdash t: A$ in $\HA+\NEM$. Then $t\in\sn$.
\end{corollary}

\textit{Proof}.
Assume $\Gamma=x_1: {A_1},\ldots,x_n:{A_n}, \Delta$ ($\Delta$ not containing declarations of proof-term variables). By \cruno, one has $x_{i}\red A_{i}$, for $i=1,\ldots, n$. 
From Theorem \ref{Adequacy Theorem}, we derive that $t\red A$. From \cruno,  we conclude that $t\in\sn$.\\

The strong normalization of $\HA+\NEM$ is readily turned into a strong normalization result for $\HA+\EM_{1}$, since the reduction $\mapsto$ can be simulated by $\redn$. 

\begin{corollary}[Strong Normalization for $\HA+\EM_{1}$]\label{theorem-snNEM} Suppose $\Gamma\vdash t: A$ in $\HA+\EM_{1}$. Then $t\in\sn$.
\end{corollary}

\textit{Proof}.
 By Proposition \ref{proposition-preservation}, any infinite reduction $t=t_{1}\mapsto t_{2}\mapsto \ldots \mapsto t_{n} \mapsto \ldots $ in $\HA+\EM_{1}$  gives rise to an infinite reduction $\trans{t}=\trans{t}_{1} \redn^{+} \trans{t}_{2}\redn^{+} \ldots\redn^{+} \trans{t}_{n}\redn^{+} \ldots$ in $\HA+\NEM$.  By the strong normalization Corollary \ref{theorem-snNEM} for $\HA+\NEM$ and since clearly $\Gamma\vdash \trans{t}: A$, infinite reductions of the  latter kind cannot occur; thus neither of the former. 

\section{Conclusions}

So far, the main application of  the strong normalization of the system $\HA+\NEM$ under the relation $\redn$ is the proof of the strong normalization of $\HA+\EM_{1}$ under $\mapsto$. In the future, when attempting to prove the strong normalization of an exception-based Curry-Howard correspondence for $\mathsf{PA}$, our method could reveal itself to be crucial as a first smooth tool to prove strong normalization in a considerably more complicate setting. While it is far from obvious how to extend the realizability in $\cite{ABB}$, the non-deterministic method does not need semantical insights and seems more flexible. 

 We also remark that the relation $\redn$ is not at all just a meaningless trick used to  simplify the realizability proof in \cite{ABB}, and is interesting in its own. For example, it expresses all the possible meaningful computations of witnesses that can be extracted from a proof in $\HA+\EM_{1}$. This holds because the reduction $\E{}{u}{v}\redn v$ combined with the reduction $\Wit{P}{\alpha}\redn (n, \True)$, allows to consider all the possible witnesses for the hypothesis $\exists \alpha^{\Nat}\emp{}$. Thus the reduction $\redn$ is useful to study the non-determinism of classical proofs and also shows that indeed many more effective reductions can be considered other than those coded in the relation $\mapsto$, and   strong normalization will nevertheless be preserved. For example, in order to reduce $\E{}{u}{v}\redn v$ is not necessary to wait for a witness raised by an exception in $u$, and the witness can be obtained from anywhere point all-over the surrounding context or in whatever else way. As a consequence,  the reduction $\redn$ may be allowed to have a companion \emph{state} as in Interactive realizability, where all the known witnesses may be collected: the reduction step $\E{}{u}{v}\redn v$ can be done whenever  in the state there is a witness for replacing in a logically correct way the appropriate hypothesis $\Wit{P}{\alpha}$ occurring in $v$. This allows the programs to be much more efficient, because re-computation of witness may always be avoided. In the relation $\mapsto$ this optimization is even not legal, because a term $\E{a}{u}{v}$ must always wait for an exception in order to reduce to $v[a:=n]$ for some $n$, even if that $n$ has already been computed when reducing another instance of the program $\E{a}{u}{v}$ in the past of the computation.

 \providecommand{\urlalt}[2]{\href{#1}{#2}}
\providecommand{\doi}[1]{doi:\urlalt{http://dx.doi.org/#1}{#1}}


\begin{thebibliography}{[1]} 

{
 


\bibitem{ABF} 
 Aschieri, F.,  Berardi, S: \emph{A New Use of Friedman's Translation: Interactive Realizability}, in: Logic, Construction, Computation, Berger et al. eds, Ontos-Verlag Series in Mathematical Logic, 2012.

\bibitem{AschieriCSL}  Aschieri, F.: \emph{Interactive Realizability for Classical Peano Arithmetic with Skolem Axioms}. Proceedings of Computer Science Logic 2012, 
Leibniz International Proceedings in Informatics, vol. 16, 2012. \doi{10.4230/LIPIcs.CSL.2012.31}
\bibitem{ABB} 
 Aschieri, F., Berardi, S., Birolo, G.: \emph{Realizability and Strong Normalization for a Curry-Howard Interpretation of HA + EM1}. Proceedings of CSL 2013.

\bibitem{AschieriZorzi} Aschieri, F., Zorzi, M.:  \emph{Non-Determinism, Non-Termination and the Strong Normalization of System T}, Proceedings of TLCA 2013: 31--47. \doi{10.1007/978-3-642-38946-7\_5} 


\bibitem{AvigadR} Avigad, J.: \emph{A realizability Interpretation for Classical Arithmetic}, in Buss et al. eds., Logic Colloquium '98, Lecture Notes in Logic 13, AK Peters, 57-90, 2000.

\bibitem{Birolo}
 Birolo, G.: \emph{Interactive Realizability, Monads and Witness Extraction}, Ph.D. thesis, June, 2013, Universit\`{a} di Torino. http://arxiv.org/abs/1304.4091

\bibitem{BerardiLiguoroMonadi} Berardi, S., de' Liguoro, U.:
\emph{Interactive Realizers. A New Approach to Program Extraction from Nonconstructive Proofs}, ACM Transactions on Computational Logic, 2012. \doi{10.1145/2159531.2159533}



\bibitem{BB2} Barbanera, F., Berardi, S.: \emph{Extracting Constructive Content from Classical Logic via Control-like Reductions}, TLCA 1993, 45-59. \doi{10.1007/BFb0037097}

\bibitem{BB1} Barbanera, F., Berardi, S.: \emph{A Constructive Valuation Semantics for Classical Logic}, Notre Dame Journal of Formal Logic, {\bfseries 37(03)} (1996). 


\bibitem{BB3} F. Barbanera, S. Berardi,
\emph{A Symmetric Lambda-Calculus for Classical Program Extraction}, Information and Computation {\bfseries 125(02)} (1996). \doi{10.1006/inco.1996.0025}

\bibitem {MarghiUgo}
Dal Lago, U., Zorzi, M.:
\emph{Probabilistic Operational Semantics for the Lambda Calculus}.
RAIRO-ITA  {\bfseries 46(03)} (2012) 413--450. \doi{10.1051/ita/2012012}

\bibitem {David1}
David, R., Nour, K:
 \emph{A short proof of the strong normalization of the simply typed lambda-mu-calculus}.
Schedae Informaticae {\bfseries 12} (2003) 27--34. 



\bibitem {DeLPip1}
de' Liguoro, U., Piperno, A.:
\emph{Non-Deterministic Extensions of Untyped Lambda-Calculus}.
Information and Computation {\bfseries 122} (1995) 149--177. \doi{10.1006/inco.1995.1145}

\bibitem {Girard}
Girard, J.-Y., Lafont, Y., Taylor, P.:
\emph{Proofs and Types}.
Cambridge University Press
{\bfseries } (1989). 

\bibitem{Herbelin} Herbelin, H.: \emph{An Intuitionistic Logic that Proves Markov's Principle}, Proceedings of LICS 2010: 50-56. \doi{10.1109/LICS.2010.49}

\bibitem{deGrooteex} de Groote, P.: \emph{A Simple Calculus of Exception Handling}, Proceedings of TLCA 1995: 201--215. \doi{10.1007/BFb0014054}

\bibitem{deGroote} de Groote, P.: \emph{Strong Normalization for Classical Natural Deduction with Disjunction}, Proceedings of TLCA 2001: 182--196. \doi{10.1007/3-540-45413-6\_17}

\bibitem{Griffin} Griffin, T.: \emph{A Formulae-as-Type Notion of Control}, Proceedings of POPL, 1990. \doi{10.1145/96709.96714}





\bibitem{Krivine1}
Krivine, J.-L.:
\emph{Lambda-calcul types et mod\`eles}, Studies in Logic and Foundations of Mathematics {\bfseries} (1990) 1--176.  Masson, Paris.

\bibitem{Krivine2}
Krivine, J.-L.: \emph{Classical Realizability}. In Interactive models of computation and program behavior. Panoramas et synth\`eses {\bfseries 27} (2009) 197--229.  Soci\'et\'e Math\'ematique de France.
\bibitem{Krivine3} Krivine, J.-L.: \emph{Realizability Algebras II: new models of ZF + DC}, Logical  Methods in Computer Science, {\bfseries 8(1)} (2012). \doi{10.2168/LMCS-8(1:10)2012}




\bibitem{Nour} Nour, K., Saber, K.: \emph{A Semantical Proof of the Strong Normalization Theorem for full Propositional Classical Natural Deduction}, Archive of Mathematical Logic {\bfseries45(3)} (2006), 357--364. \doi{10.1007/s00153-005-0314-y}


\bibitem{Parigot1} Parigot, M.: \emph{Classical Proofs as Programs}, Kurt G\"odel Colloquium 1993, 263--276. \doi{10.1007/BFb0022575}

\bibitem{Parigot2} Parigot, M.: \emph{Proofs of Strong Normalization for Second Order Classical Natural Deduction}, Journal of Symbolic Logic {\bfseries 62(4)} (1997) 1461-1479. 



\bibitem{Prawitznatural}
Prawitz, D.: \emph{Natural Deduction, A Proof-Theoretical Study}, Almqvist \& Wiksell, Stockholm, 1965. 

\bibitem{Prawitz}
Prawitz, D.: \emph{Ideas and Results in Proof Theory}. In Proceedings of the Second Scandinavian Logic Symposiuum (1971). 



 \bibitem{Sorensen}M. H. Sorensen, P. Urzyczyn, \emph{Lectures on the Curry-Howard isomorphism}, Studies in Logic and the Foundations of Mathematics, vol. 149, Elsevier, 2006.

 \bibitem{Troesltra} Troelstra, A. S.: \emph{Realizability}. In: Barwise ed. Handbook of Proof Theory, Studies in Logic and Foundations of Mathematics, 1998, vol. 137, ch. 6, 407--473.


}
 \end{thebibliography}
\end{document}